\documentclass[sigconf]{acmart}

\acmBooktitle{Proceedings of the CHI 2026 Workshop: Ethics at the Front-End: Responsible User-Facing Design for AI Systems (CHI '26), April 13--17, 2026, Barcelona, Spain}
\acmDOI{} 

\makeatletter
\patchcmd{\@mkbibcitation}
  {ACM, New York, NY, USA\@article@string\unskip, \ref{TotPages}~\@pages@word.}
  {}
  {}{}
\makeatother

\usepackage{url}

\AtBeginDocument{%
  }

\setcopyright{acmlicensed}
\copyrightyear{2026}
\acmYear{2026}

\acmConference[CHI'26]{}{April 13--17,
  2026}{Barcelona, Spain}
\acmISBN{}

\usepackage{url}
\begin{document}

\title{Strategies for Designing Responsibly within a Capitalist Enterprise}


\author{Shixian Xie}
\email{shixianx@andrew.cmu.edu}
\affiliation{%
  \institution{Carnegie Mellon University}
  \city{Pittsburgh}
  \state{Pennsylvania}
  \country{USA}
}

\author{Motahhare Eslami}
\email{meslami@andrew.cmu.edu}
\affiliation{%
  \institution{Carnegie Mellon University}
  \city{Pittsburgh}
  \state{Pennsylvania}
  \country{USA}
}

\author{John Zimmerman}
\email{johnz@cs.cmu.edu}
\affiliation{%
  \institution{Carnegie Mellon University}
  \city{Pittsburgh}
  \state{Pennsylvania}
  \country{USA}
}

\renewcommand{\shortauthors}{Xie, Eslami, Zimmerman.}


\begin{abstract}
Despite significant advances in responsible AI research, industry adoption remains limited, leaving many HCI contributions underutilized in practice. This position paper argues that current research often fails to account for the fundamental need for capitalist enterprises to create value. To achieve immediate real-world impact, responsible AI research must explore how to design responsibly \textit{within} capitalism. We call for a move beyond the dichotomy of "ethics vs. business" toward a more productive framing of "ethics and business." We propose ideation as a practical design strategy for generating ethically preferable alternatives that also meet business objectives. By aligning ethics with enterprise realities, we expand the space of responsible design that can actually be built.
\end{abstract}




\setcopyright{acmcopyright}

\maketitle

\section{Introduction}
AI systems are increasingly shaping people’s lives, from determining what information we see to mediating access to housing \cite{kuo2023understanding}, healthcare \cite{nazar2021systematic}, and employment \cite{fabris2025fairness}. With this influence has come a growing awareness of AI’s harms. In response, a vibrant research community emerged around designing responsible AI systems. Responsible AI and HCI researchers made significant contributions to minimizing AI’s risks, including work on identifying and addressing algorithmic bias \cite{shen2021everyday,solyst2023potential}, as well as exposing and challenging user manipulation design practices, such as dark patterns in user interfaces \cite{gray2018dark,mildner2023engaging}.

Despite this progress, responsible AI approaches remain underutilized in enterprise adoption. Companies continue to deploy systems that optimize for revenue and engagement—often through mechanisms that HCI researchers thoroughly critiqued. Without enterprise adoption, the benefits of the responsible AI research go unrealized. This lack of adoption prompts a critical question: how can we increase the practical relevance and real-world adoption of our community’s work?

This position paper argues that designing responsibly within a capitalist enterprise requires a better strategy: ideation—the process of generating multiple design alternatives that are both ethically preferable and financially viable. Current responsible design proposals often overlook the fundamental needs for enterprises to create value. As a result, researchers frequently propose solutions that, while ethically sound, offer little financial incentive for companies to adopt them. If we want ethical AI to scale, we must learn to design interventions that align ethical commitments with business imperatives.

\section{Our position: Ethics \textit{and} Business}
Despite growing concern for AI’s harms, our research community continues to produce design proposals that are difficult for most businesses to adopt. Many of these ideas are incompatible with the financial models that sustain product development, making them unappealing or even infeasible for capitalist enterprises. If we want responsible AI to have an immediate real-world impact, we must learn how to design \textit{within} capitalism. This means acknowledging the motivations, constraints, and pressures that shape design decisions in real-world capitalist enterprise environments.

A clear example of this tension is the widespread use of dark patterns, an interface design practice that exploits user psychology to increase engagement, retention, or monetization. Companies adopt dark patterns because they directly support key financial metrics. Within capitalist enterprises, these financial metrics define success. Yet these design choices often conflict with ethical principles, such as autonomy, informed consent, and user well-being. In response, much of the current responsible AI literature turned to policy and regulation to address those harms. While this work is important, policy reform is inherently slow: lawmaking takes time, and enforcement takes even longer. Meanwhile, manipulative systems continue to be deployed at scale, far outpacing regulatory responses. Dark patterns offer a striking example – despite years of critical research and increasing regulatory interest, they remain common in consumer interfaces today. These examples illustrate that legal strategies alone are unlikely to respond quickly enough to prevent ongoing harm. 

We argue for a complementary strategy to design responsibility within a capitalist enterprise. Researchers need to take the financial logic into account when proposing solutions to unethical design practices. Effective solutions must attend to both the harm and the financial incentive structures that drive it. We need to stop treating business models as separate from design practices, as if ethics can succeed on moral clarity alone. Business models and design practices are not separate domains; they are deeply intertwined. \textit{If we continue to ignore business value, responsible design will continue to be ignored in practice.}

One way to achieve the balance between ethical design and enterprise value is through ideation \cite{jonson2005design} – the process of generating multiple design alternatives before committing to a single design decision. This process allows teams to surface design options that have equal business value and lower responsible AI risks. Without ideation, teams have only one idea to work on, missing the opportunity to identify design alternatives that address both business needs and ethics simultaneously.

No company wants to appear in the news headlines for responsible AI issues. We recommend that when faced with a responsible AI challenge, teams ideate a range of possible alternatives to explore what other innovations the enterprise might develop, other than this problematic one. The goal here is to reveal an opportunity of equal value, but without the harm. Generating alternatives avoids the binary view of choosing between “ethical design or business value” and instead promotes solutions that are both. When teams have more than one idea to work on, it reframes the question from “Is this design ethical?” to “Which of these ethically-informed options will maximize AI benefits and minimize harms?” In doing so, we surface designs that enterprises might actually choose to build—and that users might actually benefit from.

\section{Call to Action}
We argue that achieving immediate real-world impact in responsible AI design requires researchers to pursue ethical principles and business value simultaneously. Responsible AI cannot scale if it only critiques harmful systems without also generating deployable alternatives within capitalist enterprises. We call on the field to recognize and support research that advances both ethical outcomes and enterprise viability. This is not a call to deprioritize existing responsible AI work. However, without parallel investment in research that operates within real-world economic systems, responsible AI will remain under-deployed in practice.

Call To Action 1. Advance Ideation as a Core Responsible Design Strategy. HCI researchers should develop methods, toolkits, and evaluation frameworks that support generating multiple ethically preferable and enterprise-viable design alternatives. Establishing ideation as a responsible design approach will expand the set of ethical systems that can realistically move into enterprises.

Call To Action 2. Value Responsible Design Research that Demonstrates Enterprise Feasibility. Conference reviewers, program committees, and subcommittees should explicitly value research that integrates ethical design outcomes with business value and deployment constraints. Clear signaling through calls for papers, reviewer guidelines, and awards will legitimize this research direction and accelerate its growth.

Call To Action 3. Welcome Business Perspectives into the Field. HCI must actively welcome business researchers. Expanding interdisciplinary collaboration will increase the likelihood that responsible design research produces systems that are attractive to a capitalist enterprise. 

\section{Conclusion}
Responsible AI design must be grounded in the economic realities where it is developed—particularly within capitalist enterprise environments. When detached from these contexts, responsible design solutions risk being dismissed as impractical or irrelevant. We propose ideation as a strategy for generating responsible solutions that are both ethically sound and commercially viable. By accounting for business imperatives, responsible design is more likely to achieve immediate real-world impact.

\bibliographystyle{ACM-Reference-Format}
\bibliography{References}

\end{document}